\documentclass[useAMS,usenatbib]{mn2e}
\usepackage{epsfig}
\usepackage{color}

\title[DM scaling relations in intermediate $z$ haloes]{Dark matter scaling relations in intermediate $z$ haloes}
\author[V.F. Cardone \& C. Tortora]{V.F. Cardone$^{1,2}$ and C. Tortora$^{3}$\\
$^1$ Dipartimento di Scienze e Tecnologie dell' Ambiente e del Territorio, Universit\`{a} degli Studi del Molise, \\
Contrada Fonte Lappone, 86090\,-\,Pesche (IS), Italy \\
$^2$ Dipartimento di Scienze Fisiche, Universit\`{a} degli Studi di Napoli "Federico II",
Complesso Universitario \\ di Monte Sant'Angelo, Edificio N, via Cinthia, 80126 - Napoli, Italy \\
$^3$ Universit$\ddot{a}$t Z$\ddot{u}$rich, Institut f$\ddot{u}$r
Theoretische Physik, Winterthurerstrasse 190, CH-8057, Z$\ddot{u}$rich, Switzerland\\}

\date{Accepted xxx, Received yyy, in original form zzz}

\begin{document}
\maketitle

\begin{abstract}

We investigate scaling relations between the dark matter (DM) halo
model parameters for a sample of intermediate redshift
early\,-\,type galaxies (ETGs) resorting to a combined analysis of
Einstein radii and aperture velocity dispersions. Modeling the
dark halo with a Navarro\,-\,Frenk\,-\,White profile and assuming
a Salpeter initial mass function (IMF) to estimate stellar masses,
we find that the column density ${\cal{S}}$ and the Newtonian
acceleration within the halo characteristic radius $r_s$ and
effective radius $R_{eff}$ are not universal quantities, but
correlate with the luminosity $L_V$, the stellar mass $M_{\star}$
and the halo mass $M_{200}$, contrary to recent claims in the literature. We finally discuss a tight
correlation among the DM mass $M_{DM}(R_{eff})$ within the
effective radius $R_{eff}$, the stellar mass $M_{\star}(R_{eff})$
and $R_{eff}$ itself. The slopes of the scaling relations discussed
here strongly depend, however, on the DM halo model and the IMF
adopted so that these ingredients have to be better constrained in
order to draw definitive conclusions on the DM scaling relations
for ETGs.

\end{abstract}

\begin{keywords}
dark matter -- galaxies\,: kinematic and dynamics -- galaxies\,: elliptical and lenticulars, CD
\end{keywords}

\section{Introduction}

The current cosmological paradigm, the concordance $\Lambda$CDM
model, relies on two main components, namely dark energy (e.g.
\citealt{CPT92}) and dark matter. Although in excellent agreement
with all the cosmological probes (\citealt{WMAP5,P09,SNeIaSDSS}),
the $\Lambda$CDM model is nevertheless afflicted by serious
problems on galactic scales. In this framework, the formation of
virialized DM haloes from the initial tiny density perturbations
is followed at later stages through numerical N\,-\,body
simulations (\citealt{B98}). It became apparent that the
spherically averaged density profile, $\rho_{DM}(r)$, of DM haloes
is independent of the halo mass (\citealt{NFW97}) and well
described by a double power\,-\,law relation with $\rho_{DM}
\propto r^{-3}$ in the outer regions and $\rho_{DM} \propto
r^{-\alpha}$ with $\alpha > 0$ centrally. On the contrary,
observations of spiral galaxies seem to definitely point towards
cored models, i.e. $\alpha = 0$ at the centre (\citealt{deB09}).
Understanding whether such a discrepancy is due to some physical
process not correctly modeled in simulations or to a failure of
the CDM paradigm is still a hotly debated issue. As a valuable tool to
address this problem, one can look for scaling relations among DM
halo parameters and stellar quantities in order to better
constrain the formation scenario and the DM properties. Recently,
much work has been dedicated to this issue. Using a sample of
local ETGs, \citet[T09 hereafter]{T09} have found that DM is the
main driver of the Fundamental plane (FP) tilt (see also
\citealt{Cappellari+06}, \citealt{Bolton+07}, \citealt{HB09}, \citealt{GF09}, \citealt{SlacsX}) and
that the average spherical DM density is a decreasing function of
stellar mass (see also \citealt{Thomas+09}). Based on data from rotation curves of $\sim 1000$ spiral galaxies, the mass models of
individual dwarf and spiral galaxies and the weak lensing signal
of elliptical and spirals, \citet[D09]{D09} and \citet[G09]{G09}
have found strong evidence for the constancy of the central DM
column density over 12 orders of magnitude in luminosity.
\citet[NRT10]{NRT10} have shown that, on average, the projected
density of local ETGs within effective radius is systematically
higher than the same quantity for spiral and dwarf galaxies,
pointing to a systematic increase with halo mass as suggested by
\citet[B09]{B09}, who have extended the samples
analyzed above to both group and cluster scale systems.

In order to try to discriminate between these contrasting results,
we present here an analysis of the DM scaling relations for a
sample of ETGs at intermediate redshift ($\langle z \rangle \simeq
0.2$) using lensing and velocity dispersion data to constrain
their parameters. The mass models, the data and the fitting
procedure are described in \S\ref{sec:sec1}. In \S\ref{sec:sec2}
we describe the main results, while \S\ref{sec:sec3} is devoted to
a brief review of the results and conclusions.

\section{Estimating mass quantities}\label{sec:sec1}

As a preliminary mandatory step, we need to determine the
quantities involved in the above scaling relations. To this end,
one has first to choose a model for the stellar and DM components
and then fit the observational data in order to infer the
quantities of interest from the constrained model.

\subsection{Stellar and DM profiles}

Motivated by the well known result that the surface brightness
profiles of ETGs are well fitted by the S$\rm\acute{e}$rsic (1968) law, we
describe the stellar component with the \citet[PS hereafter]{PS97}
profile (see also \citealt{Mar01} for further details). The choice
of the DM halo model is quite controversial. Rotation curves of $z
= 0$ spiral galaxies are better fitted by cored models
(\citealt{deB09}), but we are here considering ETGs at
intermediate $z$ so that it is not straightforward to extend these
results to our case. In order to explore the impact of the DM halo
profile, we therefore adopt both a \cite{B95} model with

\begin{equation}
\rho_B(r) = \frac{\rho_B r_B^3}{(r + r_B) (r^2 + r_B^2)} \ ,
\label{eq: rhobur}
\end{equation}
and an NFW (\citealt{NFW97}) profile with

\begin{equation}
\rho_{NFW}(r) = \frac{\rho_s r_s^3}{r (r + r_s)^2} \ .
\label{eq: rhonfw}
\end{equation}
Both the 3\,-\,dimensional and projected masses $M(r)$ and
$M_{proj}(R)$ can be analytically evaluated, with $M_{proj}(R)$
given in \cite{PF03} and \cite{B96} for the Burkert and NFW
models, respectively.

In order to constrain the model parameters, we rely on the
estimate of the projected mass $M_{E}=M_{proj}(R_E)$ inferred by
the measurement of the Einstein radius $R_{E}$ in a lens system.
While lensing probes the mass projected along the line of sight,
the aperture velocity dispersion $\sigma_{ap}$ \citep{ML05}
provides complementary information on the internal dynamics thus
strengthening the constraints.

\subsection{Data and fitting procedure}

We make use of the sample of 85 lenses collected by the Sloan Lens
ACS (SLACS) survey (\citealt{Slacs}) and first select only ETGs
with available values of both the velocity dispersion
$\sigma_{ap}$ (measured within an aperture of $R_{\rm ap}= 1.5''$)
and the Einstein radius $R_E$, thus ending up with a dataset
containing 59 objects. For each lens, we follow \cite{Slacs}
setting the S$\rm\acute{e}$rsic index $n = 4$ and the effective
radius $R_{eff}$ and total luminosity $L_V$ to the values inferred
from the $V$\,-\,band photometry. The SLACS collaboration has also
provided an estimate of the total stellar mass (their Table 4) from which we use both \cite{Chabrier01} and
\cite{Salpeter55} IMFs to investigate the effect of the IMF on the
scaling relations.

Before fitting the model to the data, it is worth noting that, for
both the Burkert and NFW models, the halo parameters are different
from one lens to another. Since we have only two observed
quantities, namely $(M_E, \sigma_{ap})$ for each lens, determining
$(\rho_X, r_X)$ (with $X = B$ or $s$ for the Burkert and NFW
models, respectively) on a case\,-\,by\,-\,case basis would give
us very weak constraints. We therefore bin the galaxies in 10
equally populated luminosity bins (the last one actually containing 5 objects) and resort to a different parametrization
using quantities that it can be reasonably assumed to be the same for
all the lenses in the same bin\footnote{Note that this procedure
allows us to fit for these two alternative parameters on a
bin\,-\,by\,-\,bin basis thus having $2 {\cal{N}}_{bin} - 2$
degrees of freedom with ${\cal{N}}_{bin}$ the number of lenses in
the bin.}. As one of the fitting parameters, we choose the virial
$M/L$ ratio, $\Upsilon_{vir} = M_{vir}/L_{V}$ with $M_{vir}$ the
DM halo mass at the virial radius\footnote{We define $R_{vir}$ as
the radius where the mean density $M_{vir}/(4/3) \pi R_{vir}^3$
equals $\Delta_c(z) \bar{\rho}_M(z)$ with $\Delta_c(z)$ as in
Bryan \& Norman (1998) and $\bar{\rho}_M$ the mean matter density
at $z$.} $R_{vir}$. Should $\Upsilon_{vir}$ depend on $L_V$, our luminosity
bins are quite narrow so that any change in $\Upsilon_{vir}$ should be so small
that can be safely neglected. We then use $\log{\eta_X}$ with $\eta_X =
r_X/R_{eff}$ as the second parameter.

Note that, although $(\Upsilon_{vir}, \log{\eta_X})$ are the same
for all the galaxies in a bin, $(M_{vir},r_X)$ still change from one lens to
another thus allowing us to estimate all the quantities of interest on a
lens\,-\,by\,-\,lens basis. In order to constrain
$(\Upsilon_{vir}, \log{\eta_X})$, we maximize a suitable
likelihood function, composed of two terms, the first (second) one
referring to the lensing (dynamics) constraints (see \citealt{C09}
for further details). In order to efficiently explore the
parameter space, we use a Markov Chain Monte Carlo (MCMC)
algorithm running chains with 100000 points reduced to more than
3000 after burn in cut and thinning. For a given galaxy, we
compute the quantities of interest for each point in the chains
and then use Bayesian statistics to infer median values and $68\%$
confidence intervals. We follow \cite{Dago04} to correct for the
asymmetric errors in our estimates.

\section{Results}\label{sec:sec2}

Before investigating scaling relations, it is worth checking
whether our PS  + DM model works in fitting the lens data.

\begin{table*}
\caption{Constraints on halo parameters for the Burkert models with Chabrier and Salpeter IMF.
The median luminosity from the galaxies in each bin is reported in column 1, while the adjacent columns give
the maximum likelihood parameter $(\Upsilon_{vir}, \log{\eta_s})_{ML}$, the value of $-2 \ln{{\cal{L}}}$ at the maximum, median value and $68$ per cent confidence interval for the DM parameters $(\Upsilon_{vir}, \log{\eta_s})$ for each model.}
\label{tab:_tab1a}
\begin{center}
\begin{tabular}{|c|c|c|c|c|c|c|c|c|}
\hline
Bin & \multicolumn{4}{|c|}{Chabrier IMF} & \multicolumn{4}{|c|}{Salpeter IMF} \\
\hline
$\log{L_V}$ &  $(\Upsilon_{vir}, \log{\eta_s})_{ML}$ & $-2 \ln{{\cal{L}}_{max}}$ & $(\Upsilon_{vir})_{-1\sigma}^{+1\sigma}$ & $(\log{\eta_s})_{-1\sigma}^{+1\sigma}$ &  $(\Upsilon_{vir}, \log{\eta_s})_{ML}$ & $-2 \ln{{\cal{L}}_{max}}$ & $(\Upsilon_{vir})_{-1\sigma}^{+1\sigma}$ & $(\log{\eta_s})_{-1\sigma}^{+1\sigma}$ \\
\hline \hline
10.72 & $(15, -0.61)$ & 353.179 & $26_{-12}^{+22}$ & $-0.38_{-0.24}^{+0.21}$ & $(15, -0.80)$ & 351.645 & $19_{-11}^{+32}$ & $-0.24_{-0.32}^{+0.19}$ \\

10.82 & $(53, -0.18)$ & 362.797 & $46_{-32}^{+60}$ & $-0.22_{-0.22}^{+0.22}$ & $(35, -0.15)$ & 364.679 & $41_{-23}^{+30}$ & $-0.09_{-0.25}^{+0.15}$ \\

10.85 & $(9, -0.81)$ & 367.780 & $14_{-6}^{+11}$ & $-0.62_{-0.21}^{+0.22}$ & $(15, -0.80)$ & 364.975 & $14_{-8}^{+42}$ & $-0.36_{-0.31}^{+0.38}$ \\

10.95 & $(32, -0.37)$ & 364.059 & $43_{-21}^{+48}$ & $-0.29_{-0.21}^{+0.20}$ & $(17, -0.41)$ & 364.379 & $29_{-16}^{+33}$ & $-0.25_{-0.25}^{+0.21}$ \\

11.00 & $(27, -0.31)$ & 366.641 & $36_{-16}^{+19}$ & $-0.21_{-0.20}^{+0.17}$ & $(15, -0.26)$ & 366.721 & $37_{-23}^{+54}$ & $0.02_{-0.32}^{+0.23}$ \\

11.07 & $(10, -0.80)$ & 375.608 & $13_{-5}^{+8}$ & $-0.26_{-0.23}^{+0.18}$ & $(10, -0.87)$ & 372.049 & $11_{-5}^{+22}$ & $-0.54_{-0.21}^{+0.28}$ \\

11.12 & $(24, -0.41)$ & 370.943 & $38_{-18}^{+34}$ & $-0.26_{-0.23}^{+0.18}$ & $(14, -0.40)$ & 370.400 & $40_{-24}^{+60}$ & $-0.07_{-0.29}^{+0.23}$ \\

11.15 & $(20, -0.43)$ & 377.986 & $39_{-20}^{+27}$ & $-0.23_{-0.21}^{+0.19}$ & $(8, -0.58)$ & 377.817 & $18_{-10}^{+33}$ & $-0.26_{-0.28}^{+0.28}$ \\

11.27 & $(16, -0.61)$ & 388.842 & $23_{-16}^{+23}$ & $-0.48_{-0.21}^{+0.25}$ & $(10, -0.60)$ & 384.703 & $26_{-15}^{+34}$ & $-0.65_{-0.29}^{+0.35}$ \\

11.43 & $(8, -1.05)$ & 319.962 & $13_{-5}^{+15}$ & $-0.78_{-0.25}^{+0.30}$ & $(9, -1.10)$ & 319.953 & $12_{-5}^{+23}$ & $-0.65_{-0.29}^{+0.35}$ \\

\hline
\end{tabular}
\end{center}
\end{table*}

\begin{table*}
\caption{Same as Table 1, but for NFW models.}
\label{tab:_tab1b}
\begin{center}
\begin{tabular}{|c|c|c|c|c|c|c|c|c|}
\hline
Bin & \multicolumn{4}{|c|}{Chabrier IMF} & \multicolumn{4}{|c|}{Salpeter IMF} \\
\hline
$\log{L_V}$ &  $(\Upsilon_{vir}, \log{\eta_s})_{ML}$ & $-2 \ln{{\cal{L}}_{max}}$ & $(\Upsilon_{vir})_{-1\sigma}^{+1\sigma}$ & $(\log{\eta_s})_{-1\sigma}^{+1\sigma}$ &  $(\Upsilon_{vir}, \log{\eta_s})_{ML}$ & $-2 \ln{{\cal{L}}_{max}}$ & $(\Upsilon_{vir})_{-1\sigma}^{+1\sigma}$ & $(\log{\eta_s})_{-1\sigma}^{+1\sigma}$ \\
\hline \hline
10.72 & $(1697, 1.46)$ & 352.771 & $1408_{-434}^{+641}$ & $1.39_{-0.14}^{+0.14}$ & $(2074, 1.92)$ & 350.930 & $713_{-340}^{+625}$ & $1.48_{-0.23}^{+0.22}$ \\

10.82 & $(996, 1.24)$ & 364.487 & $968_{-254}^{+555}$ & $1.23_{-0.13}^{+0.18}$ & $(590, 1.30)$ & 365.630 & $547_{-228}^{+526}$ & $1.28_{-0.22}^{+0.31}$ \\

10.85 & $(544, 1.01)$ & 366.455 & $544_{-86}^{+122}$ & $1.03_{-0.08}^{+0.09}$ & $(326, 1.15)$ & 363.877 & $369_{-121}^{+380}$ & $1.23_{-0.21}^{+0.25}$ \\

10.95 & $(544, 1.00)$ & 365.776 & $551_{-89}^{+106}$ & $1.00_{-0.07}^{+0.09}$ & $(322, 1.05)$ & 365.268 & $320_{-67}^{+129}$ & $1.07_{-0.12}^{+0.17}$ \\

11.00 & $(882, 1.34)$ & 366.180 & $795_{-250}^{+486}$ & $1.31_{-0.16}^{+0.20}$ & $(534, 1.52)$ & 365.707 & $491_{-248}^{+779}$ & $1.52_{-0.31}^{+0.43}$ \\

11.07 & $(428, 0.93)$ & 377.148 & $432_{-61}^{+75}$ & $0.94_{-0.06}^{+0.07}$ & $(233, 0.96)$ & 373.223 & $235_{-50}^{+65}$ & $0.99_{-0.11}^{+0.14}$ \\

11.12 & $(511, 1.07)$ & 373.473 & $525_{-93}^{+136}$ & $1.09_{-0.10}^{+0.19}$ & $(267, 1.09)$ & 371.347 & $262_{-66}^{+91}$ & $1.10_{-0.14}^{+0.15}$ \\

11.15 & $(383, 0.99)$ & 373.643 & $391_{-59}^{+69}$ & $1.00_{-0.06}^{+0.07}$ & $(183, 0.98)$ & 374.078 & $203_{-50}^{+65}$ & $1.02_{-0.11}^{+0.18}$ \\

11.27 & $(503, 1.05)$ & 394.857 & $513_{-100}^{+120}$ & $1.05_{-0.11}^{+0.10}$ & $(274, 1.07)$ & 388.876 & $286_{-73}^{+136}$ & $1.11_{-0.15}^{+0.18}$ \\

11.43 & $(529, 0.96)$ & 322.021 & $532_{-76}^{+86}$ & $0.97_{-0.07}^{+0.06}$ & $(529, 0.96)$ & 322.015 & $532_{-77}^{+86}$ & $0.97_{-0.07}^{+0.07}$ \\

\hline
\end{tabular}
\end{center}
\end{table*}

\begin{figure*}
\centering \psfig{file=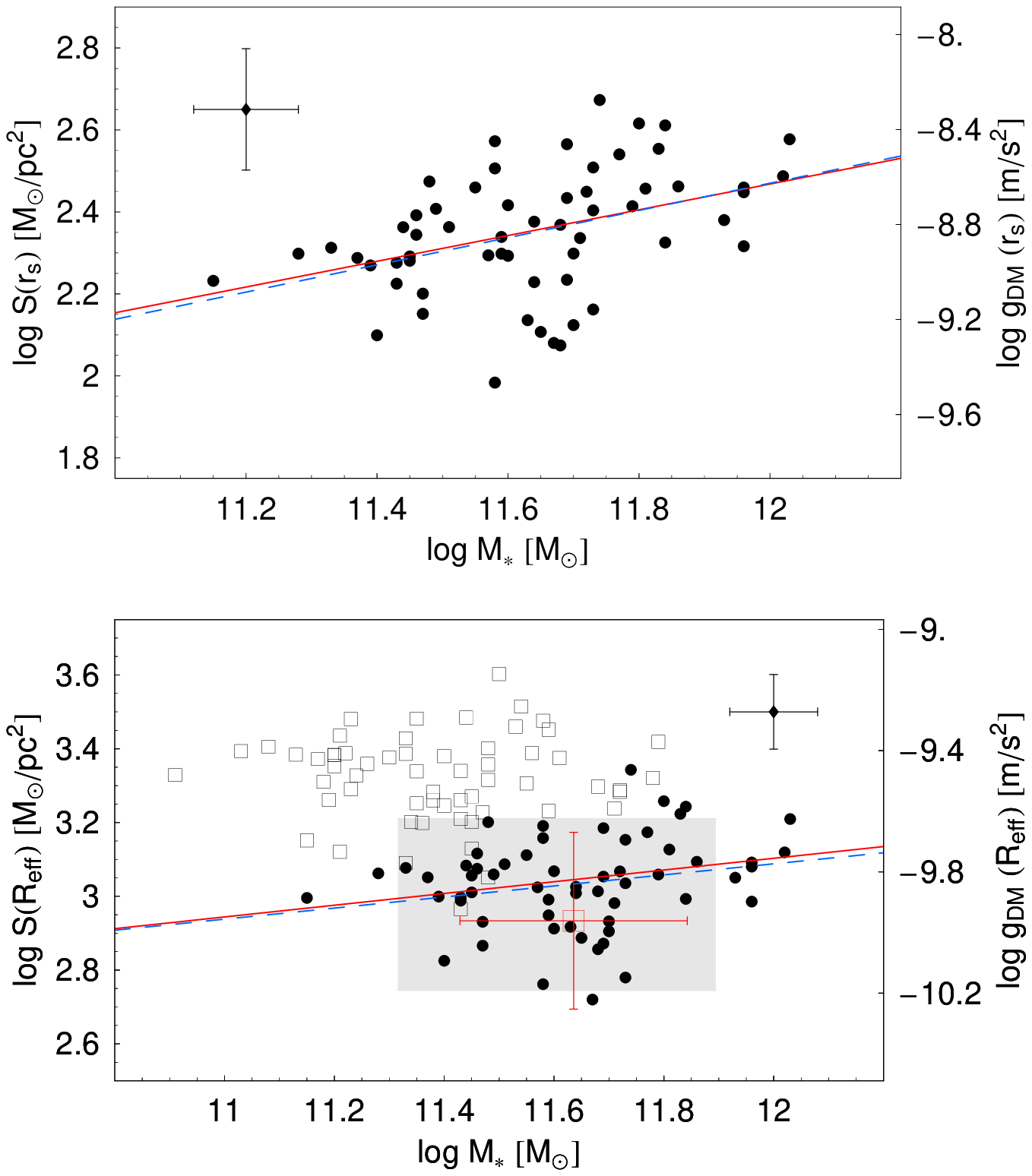,
width=0.325\textwidth} \hspace{0.4cm}
\psfig{file=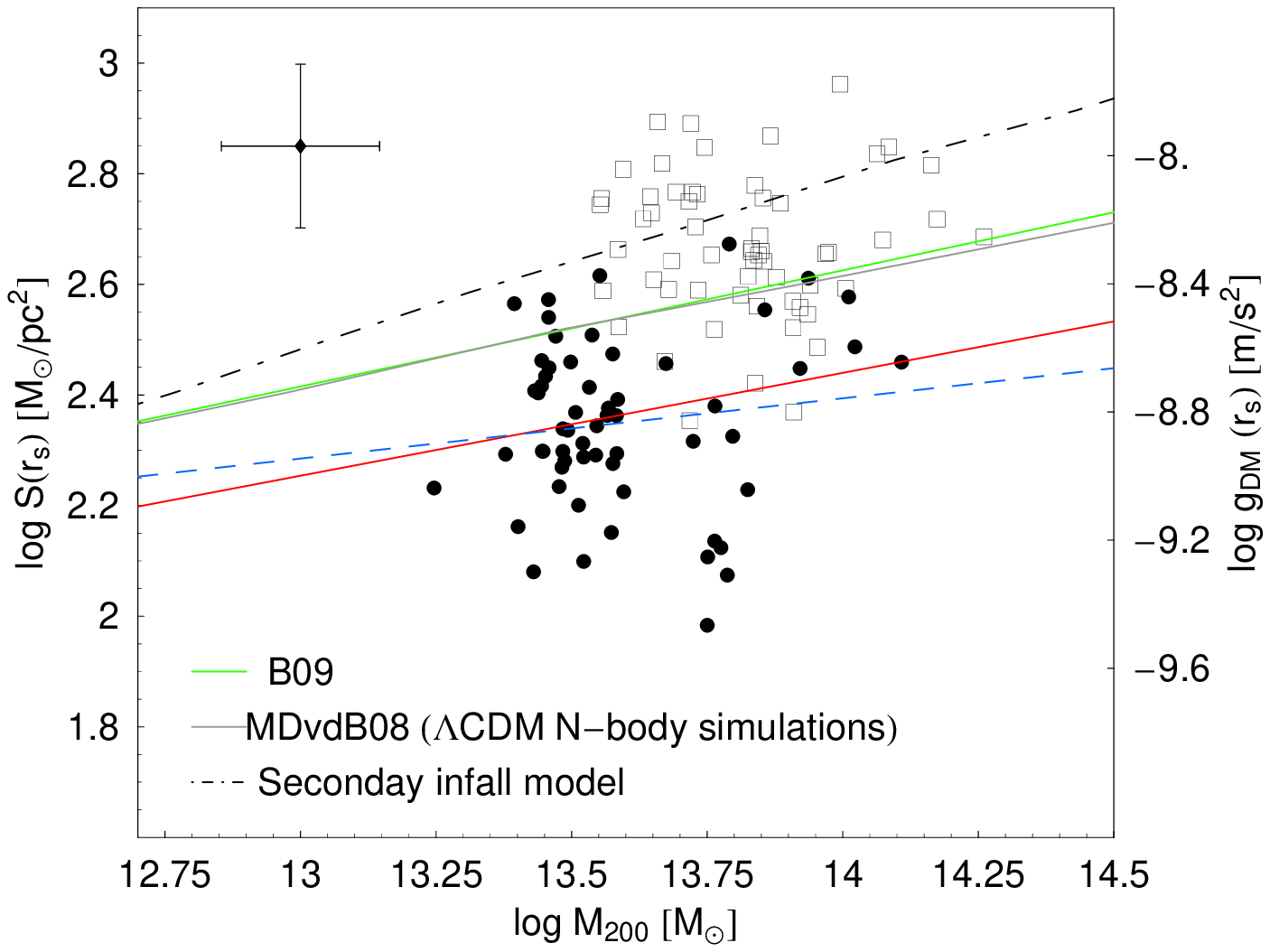, width=0.47\textwidth}
\caption{${\cal{S}}(r_s)$ and ${\cal{S}}(R_{eff})$  for the
NFW\,+\,Salpeter model as a function of stellar mass (left panels)
and ${\cal{S}}(r_s)$ vs halo mass $M_{200}$ (right panel), shown
as black points. We superimpose the best fit linear relation using
the \citet{Dago} (red solid) and direct fit (blue dashed)
methods. On the right axes is shown the equivalent acceleration
scale $g_{DM}$, defined as in the text and derived by the fit of
${\cal{S}}-g_{DM}$ relation. The black error bars set the median
data uncertainties. The results when a Chabrier IMF is used are
shown as open boxes. In the left\,-\,bottom panel, the median from
NRT10 and \citet{T+10} are shown as red symbols and gray region,
respectively (in both the cases the $1 \sigma$ scatter of the
distribution is plotted). In the right panel, we add the B09 best
fit linear relation, the secondary\,-\,infall model (Boyarsky et al. 2009b) 
prediction and the results from the $\Lambda$CDM
N\,-\,body simulation of \citet{MDvdB08}.} \label{fig: esseplots}
\end{figure*}

\subsection{The fiducial stellar\,+\,DM profile}

We have considered four PS + DM models by combining the two halo
profiles (Burkert or NFW) with the two IMFs (Chabrier or Salpeter)
adopted to set the total stellar mass. We find that all the four
models fit remarkably well the lensing and dynamical data with rms
deviations $\sim 1 \sigma$ and the lens observed values deviating no more
than $2 \sigma$ from the model ones for most of the bins. On the one
hand, such a result ensures us that the estimates of the different
quantities entering the scaling relations we are interested in are
based on empirically motivated models. On the other hand, this
same result tells us that the data we are using are unable to
discriminate among different choices. This is an outcome of the
limited radial range the data probe. Indeed, given typical values
of $R_E/R_{eff}$ and $R_{ap}/R_{eff}$, both $M_E$ and
$\sigma_{ap}$ mainly probe the inner regions of the lenses so that they
can provide only weak constraints on the behaviour of the mass
profile in the outer DM halo dominated regions.

Although statistically equivalent (as can be quantitatively
judged on the basis of the close values of ${\cal{L}}_{max}$ in Tables 1 and 2),
the four models may be ranked by examining the constraints on their parameters. Indeed,
the values in Table 1 show that the PS + Burkert models fit the data with quite low
values of $\Upsilon_{vir}$ thus leading to unexpectedly small
virial masses. Roughly averaging the maximum likelihood
$\Upsilon_{vir}$ of the different bins, we get $\langle
\Upsilon_{vir} \rangle = 22 \pm 14$ ($12 \pm 9$) $M/L_{V,\odot}$
using the Chabrier (Salpeter) IMF. On the contrary, when adopting
the NFW profile, we find $\langle \Upsilon_{vir} \rangle = 590 \pm
210$ ($360 \pm 150$) $M/L_{V,\odot}$ for a Chabrier (Salpeter)
IMF. Our results from NFW\,+\,Salpeter are qualitatively
consistent with previous estimates of $\Upsilon_{vir} \sim 200 \
\rm M/L_{V,\odot}$ obtained relying on galaxy\,-\,galaxy lensing
(\citealt{GS02, H05}) or combining strong and weak lensing and
central dynamics (\citealt{G07}). Therefore, we will consider the
NFW\,+\,Salpeter as our fiducial model\footnote{In order to put
this choice on firmer statistical grounds, we could have redefined our likelihood
by adding a prior on $\Upsilon_{vir}$ rather than the uninformative flat one we have
used here. However, since previous estimates of $\Upsilon_{vir}$ are affected by large errors
and based on model assumptions, we have preferred to avoid the risk of a fit driven by the prior
examining a posteriori the resulting values.}, retaining the other cases
just for investigating the dependence of the scaling relations on
the halo profiles and IMF adopted. This assumption is also
confirmed by recent findings which point to a Salpeter IMF when an
uncontracted NFW profile is assumed (\citealt{Treu+09}; NRT10, \citealt{A10}).

The constraints on $(\Upsilon_{vir}, \log{\eta_s})$ for the
four cases considered are summarized in Tables 1 and 2 for Burkert and NFW models respectively.
The marginalized constraints on $\Upsilon_{vir}$ show that the NFW\,+\,Salpeter model provides
reasonable values for the virial $M/L$ ratio in all bins (but the
first) thus motivating our choice as a fiducial case. The $68\%$
CL for $\Upsilon_{vir}$ are, however, quite large and
asymmetrically extended towards very large values. This can be
qualitatively explained noting that the larger is
$\Upsilon_{vir}$, the larger is $M_{vir}$ and hence $R_{vir}$. In
such a case, $R_{E}/R_{vir}$ and $R_{ap}/R_{vir}$ become quite
small such that the data are less and less able to constrain the
outer regions. As a consequence, the marginalized likelihood
function ${\cal{L}}(\Upsilon_{vir})$ has a long flat tail for
large $\Upsilon_{vir}$ thus giving rise to the reported asymmetric
errors. A similar effect also explains why the errors on
$\log{\eta_s}$ are asymmetric and still large. The wide confidence
intervals prevent us from quantifying how $\Upsilon_{vir}$ and
$\log{\eta_s}$ change from one bin to another. Excluding the first
bin, characterized by large uncertainties, we find
that both $(\Upsilon_{vir}, \log{\eta_s})$ do not show any significant trend
with $L_V$.

\subsection{DM correlations}

We now investigate the relation between the column density
${\cal{S}}(R) = M_{proj}(< R)/\pi  R^2$, luminosity, stellar and
halo mass\footnote{Note that, in  order to be consistent with B09,
we use $M_{200}$, i.e. the mass within the radius $R_{200}$, where
the mean density is 200 times the cosmological mean matter
density.} for the lenses in our sample. The main results of this
analysis are shown in Fig. \ref{fig: esseplots}. Despite the small
mass range probed and the large errors, we find that ${\cal{S}}(r_s)$
is positively correlated with $L_{V}$, $M_{\star}$ and $M_{200}$,
confirming the results in B09. The slope of the correlations may
be estimated using the \cite{Dago} fitting method, which takes
into account the errors on both variables and the intrinsic
scatter $\sigma_{int}$. Concentrating on the trend with stellar
mass, for the maximum likelihood parameters\footnote{We will refer
always to the maximum likelihood values only, but the reader must
be aware that, in a Bayesian framework, what is most important are
the marginalized values and their confidence ranges given in Table
\ref{tab: tab2}. Since the median value of the slope and the
scatter are close to the maximum likelihood ones, this choice has
no impact on the discussion.}, we get

\begin{displaymath}
\log{{\cal{S}}(r_s)} = 0.29 \log{\left ( \frac{M_{\star}}{10^{11} \ {\rm M_{\odot}}} \right )} + 2.17
\end{displaymath}
with an intrinsic scatter\footnote{The intrinsic scatter accounts for the deviations
of the single galaxies from the underlying model leading to the fitted relation.} $\sigma_{int} = 0.01$. Although the
results from a direct fit\footnote{In the direct fit, we minimize
the usual $\chi^2$ assuming that the errors on the $x$ variable
are negligible and no intrinsic scatter is present. These
simplifying assumptions do not hold for most of the scaling
relations we have considered so that we resort to the
\cite{Dago} method as a more reliable procedure. However, if
$\sigma_{int}$ is small and the error on $x$ lower than that on
$y$, the two methods converge to the same maximum likelihood
values.} are only slightly different, for completeness we will
plot the best fitted relations from both the methods (see Fig.
\ref{fig: esseplots}). Similar results are found when the column
density is fitted as a function of $L_{V}$.

\begin{table}
\caption{Marginalized constraints on the scaling relation
parameters for the correlations involving the column density ${\cal{S}}$
assuming the fiducial NFW\,+\,Salpeter model. Columns are as follows\,:
1. correlation id; 2., 3., 4. median value and $68$ confidence ranges for $(\alpha, \beta, \sigma_{int})$,
where the linear relation $\log y = \alpha \log x + \beta$ is fitted and $\sigma_{int}$ is the intrisinc scatter.
Note that, due to the fitting method, for each value of $(\alpha, \sigma_{int})$, $\beta$ is set by the condition
that the likelihood is maximized, in other words we analytically marginalize over $\beta$ when determining the maximum
likelihood parameters.} We warn the reader that, in the fit, we use the luminosity $L_V$ in units of $10^{11} \ L_{\odot}$ and
the stellar (halo) mass in units of $10^{11} (10^{12}) \ {\rm M_{\odot}}$ to reduce error covariance.
\label{tab: tab2}
\begin{center}
\begin{tabular}{|c|c|c|c|}
\hline Id &  $(\alpha)_{-1 \sigma}^{+1 \sigma}$ & $(\beta)_{-1 \sigma}^{+1 \sigma}$ & $(\sigma_{int})_{-1 \sigma}^{+1 \sigma}$ \\

\hline \hline

$\log{{\cal{S}}(r_s)}$\,-\,$\log{L_V}$ & $0.28_{-0.13}^{+0.13}$ & $2.34_{-0.01}^{+0.01}$ & $0.037_{-0.027}^{+0.043}$ \\

$\log{{\cal{S}}(r_s)}$\,-\,$\log{M_\star}$ & $0.29_{-0.15}^{+0.15}$ & $2.16_{-0.09}^{+0.10}$ & $0.037_{-0.027}^{+0.043}$ \\

$\log{{\cal{S}}(r_s)}$\,-\,$\log{M_{200}}$ & $0.14_{-0.15}^{+0.15}$ & $2.11_{-0.22}^{+0.24}$ & $0.040_{-0.029}^{+0.044}$ \\

$\log{{\cal{S}}(R_{eff})}$\,-\,$\log{M_{\star}}$ & $0.14_{-0.12}^{+0.12}$ & $2.94_{-0.08}^{+0.07}$ & $0.057_{-0.032}^{+0.039}$ \\
\hline
\end{tabular}
\end{center}
\end{table}

If we fit the same relation, but replace ${\cal{S}}(r_{s})$ with
${\cal{S}}(R_{eff})$, the trends are shallower, and the zeropoints
change too. As shown in Fig. \ref{fig: esseplots}, an error
weighted average over the galaxies in the sample gives $\langle
\log {\cal{S}}(R_{eff}) \rangle \sim 3.1$ in agreement, within the
scatter, with the median $\log{{\cal{S}}(R_{eff})} \simeq 2.9$ from
the local ETG sample of T09 and NRT10 and the results in \cite{T+10}
where a different analysis, based on an isothermal profile, is used on the same lens sample.
All these results are in agreement with a scenario where ETGs surface from the merging of
late\,-\,type systems so that, at a fixed radius, their DM content
is larger than the one for spirals and dwarves ($\log{{\cal{S}}(R_{eff})} \sim 2 \ - \ 2.5$,
see NRT10 for further details). Should we use a Chabrier IMF (thus lowering the
stellar masses by a factor $\sim 1.8$) we get a larger DM content and
${\cal{S}}(R_{eff})$ which is a constant function of $M_{\star}$
(see open boxes in Fig. \ref{fig: esseplots}).

A weaker positive correlation is found when plotting
${\cal{S}}(r_s)$ vs $M_{200}$, the maximum likelihood
fit being

\begin{displaymath}
\log{{\cal{S}}(r_{s})} = 0.16 \log{\left ( \frac{M_{200}}{10^{12} \ {\rm M_{\odot}}} \right )} + 2.11
\end{displaymath}
with $\sigma_{int} = 0.010$. Our best fit relation is shallower than the B09 one, although
the slope is consistent with their one (0.21) within the large error bars. Note that we have here
explicitly taken into account the correlated errors on ${\cal{S}}(r_s)$ and $M_{200}$, while we do not know
whether the fitting method adopted by B09 does the same. We therefore cannot exclude that the difference in slope is only
an outcome of the use of different algorithms on noisy data. Actually, our ${\cal{S}}(r_s)$
values are on average smaller than those in B09 (over the same mass range). Our results are also systematically
smaller than the estimates from the $\Lambda$CDM N\,-\,body
simulations of isolated haloes from \cite{MDvdB08} and the
predictions from the secondary infall model\footnote{Note that the predictions for
the secondary infall model of Del Popolo are actually smaller than
the B09b model ones (Del Popolo et al., in preparation).} (B09, \citealt{dP09}).
Although a wider mass range has to be probed to infer any
definitive answer, we argue that the larger values in the literature
are a consequence of neglecting the stellar component. The
agreement with B09 would improve if a Chabrier IMF was used (see
open boxes in the right panel of Fig. \ref{fig: esseplots}) and,
as discussed above, the trend would flatten out.

When discussing the results for our reference model, the observed
correlations argue against the universality of the column density
proposed in D09\footnote{Actually, what D09 refers to as universal
quantity is the product $\rho_B r_B$. It is, however, possible to
show that ${\cal{S}}_{NFW}(r_s) \simeq {\cal{S}}_{B}(1.6 r_B)$
(B09). It is then just a matter of algebra to get ${\cal{S}}_B(1.6
r_B) \simeq 1.89 \rho_B r_B$ so that the constraint $\log{(\rho_B
r_B)} = 2.17 \pm 0.20$ in D09 translates into
$\log{{\cal{S}}_B(1.6 r_B)} = 2.44 \pm 0.47$ which we use as a a
comparison value.}. It is worth investigating why we and D09 reach
completely opposite conclusions. As a first issue, we note that
D09 describe the DM halo adopting the Burkert model. Should we use
this model to infer ${\cal{S}}(1.6 r_B)$, a Chabrier IMF (which
does not strongly differ from the IMFs used in D09), and plot as a
function of luminosity to be uniform with them, the best fit
relation would have been

\begin{displaymath}
\log{{\cal{S}}(1.6 r_B)} = 0.02 \log{\left ( \frac{L_V}{10^{11} \ {\rm L_{\odot}}} \right )} + 2.65
\end{displaymath}
with $\sigma_{int} = 0.15$. In such a case, within a very good
approximation, we can assume ${\cal{S}}(1.6 r_B)$ is indeed
constant with $L_V$. An error weighted average of the sample
values gives $\langle \log{{\cal{S}}(1.6 r_B)} \rangle = 3.07 \pm
0.02 \pm 0.20$, where here and in following similar estimates, the
first error is the statistical uncertainty and the second one the
rms around the mean. Taken at face value, this estimate is
significantly larger than the D09 one (similarly to what was found
when discussing the comparison among the ${\cal{S}}(R_{eff})$ values for ETGs and spirals),
even if it is marginally within $2 \sigma_{rms}$. However, as discussed above,
the slope of the ${\cal{S}}_B(1.6 r_B)$\,-\,$L_V$ relation is
strongly model dependent. Indeed, changing to a Salpeter IMF, we
find ${\cal{S}}_B(1.6 r_B) \propto L_{V}^{0.20}$ thus arguing in
favour of non universality. Actually, the Burkert model is likely
to be unable to fit the data for ETGs with reasonable values of
$\Upsilon_{vir}$ so that one should rely on the NFW\,+\,Salpeter
results to thus conclude that the column density ${\cal{S}}$
depends on mass for intermediate redshift ETGs. Investigating the
reasons why NFW models are preferred over Burkert ones for ETGs,
while the opposite is true for spirals can provide important
constraints on galaxy formation scenarios, but it is outside our
scope here.

Motivated by the D09 findings on the universality of $\rho_B r_B$
and noting that the DM Newtonian acceleration, $g_{DM}(r) = G
M_{DM}(r)/ r^2$, is proportional to $\rho_B r_B$ for a Burkert
model, G09 have recast the D09 results in terms of a constancy of
$g_{DM}(r_B)$. Then they extended this result showing that also
the stellar acceleration, $g_{\star}(r) = G M_{\star}(r)/r^2$ at
$r_B$ is a universal quantity. Using their same Burkert model and
a Chabrier IMF, we indeed find that $g_{DM}(r_B)$ is roughly
constant with $L_V$, with an error weighted value $\langle
\log{g_{DM}(r_B)} \rangle = -8.89 \pm 0.02 \pm 0.20$, in
satisfactory agreement with the G09 result ($\log{g_{DM}(r_B)}
\simeq -8.5$). Having said that, we find that $g_{\star}(r_B)$ is
not constant, the best fit relation predicting $g_{\star}(r_B)
\propto L_V^{-0.21}$. We stress, however, that the situation is
completely reversed if we use a Salpeter IMF giving $g_{DM}(r_B)
\propto L_V^{0.20}$ and $g_{\star}(r_B) \propto L_V^{-0.01}$, so that drawing a definitive answer on which quantity is universal is
an ambiguous task. Whatever is the correct IMF, we can nevertheless
safely conclude that the DM and stellar Newtonian accelerations cannot both be universal quantities, in contrast with the claim in G09. Moreover, should the IMF be universal and intermediate between Chabrier and Salpeter, one could
argue that neither $g_{DM}(r_B)$ and $g_{\star}(r_B)$ are universal quantities.

\begin{table}
\caption{Marginalized constraints on the scaling relation parameters for the correlations $\log{g_{DM}(R_{eff})}$\,-\,$\log{L_V}$ (upper part)
and $\log{g_{\star}(R_{eff})}$\,-\,$\log{L_V}$ (lower part) with the accelerations in ${\rm m/s^2}$ and the luminosity in units of $10^{11} \ {\rm L_{\odot}}$. Columns are as follows\,: 1. model id, 2., 3., 4. median value and $68$ per cent confidence ranges for $(\alpha, \beta, \sigma_{int})$. The model ids are BSC for Burkert\,+\,Chabrier, BSS for Burkert\,+\,Salpeter, NSC for NFW\,+\,Chabrier, NSS for NFW\,+\,Salpeter. Note that, since $g_{\star}(R_{eff})$ refers to stellar quantities only, the correlations involving this quantity are independent on the halo model.}
\label{tab: tab3}
\begin{center}
\begin{tabular}{|c|c|c|c|}
\hline Model Id & $(\alpha)_{-1 \sigma}^{+1 \sigma}$ & $(\beta)_{-1 \sigma}^{+1 \sigma}$ & $(\sigma_{int})_{-1 \sigma}^{+1 \sigma}$ \\

\hline \hline

BSC & $-0.155_{-0.145}^{+0.145}$ & $-9.090_{-0.003}^{+0.003}$ & $0.175_{-0.019}^{+0.023}$ \\

BSS & $-0.004_{-0.142}^{+0.144}$ & $-9.322_{-0.002}^{+0.004}$ & $0.156_{-0.029}^{+0.033}$ \\

NSC & $0.088_{-0.111}^{+0.111}$ & $-9.526_{-0.004}^{+0.004}$ & $0.066_{-0.035}^{+0.043}$ \\

NSS & $0.245_{-0.125}^{+0.126}$ & $-9.841_{-0.006}^{+0.007}$ & $0.040_{-0.028}^{+0.043}$ \\

\hline

NSC & $-0.461_{-0.168}^{+0.170}$ & $-9.649_{-0.005}^{+0.004}$ & $0.178_{-0.022}^{+0.028}$ \\

NSS & $-0.461_{-0.168}^{+0.170}$ & $-9.403_{-0.005}^{+0.004}$ & $0.178_{-0.022}^{+0.028}$ \\

\hline

\end{tabular}
\end{center}
\end{table}

As already discussed, our fiducial model is the NFW\,+\,Salpeter so
that it is preferable to use this model when investigating whether
the Newtonian accelerations are constant or not. Moreover, rather
than using the halo characteristic radius (which refers to a
different mass content depending on the model), we will discuss
the results at $R_{eff}$ thus referring to the better constrained
inner regions. For the fiducial model, we find
$\log{g_{DM}(R_{eff})} \propto 0.26 \log{L_V}$ thus arguing
against the universality of this quantity. However, as can be seen in Table 4,
the slope of the $g_{DM}(R_{eff})$\,-\,$L_V$ correlation is strongly model
dependent with values indicating either an increasing behaviour
(for NFW models), a flat one (for Burkert\,+\,Salpeter) or a
decreasing one (for Burkert\,+\,Chabrier). These trends are
expected from the analysis of column density made above.
On the contrary, the scaling of $g_{\star}(R_{eff})$ does not depend on the halo model
because its value is estimated from stellar quantities only, while the effect of the IMF is simply a
systematic rescaling due to the higher stellar masses for the Salpeter case. Although the slope
in Table 4 points towards a significative decrease with the luminosity, it is worth stressing that
assuming a constant value $\langle \log{g_{\star}(R_{eff})} \rangle = -9.403 \pm 0.01 \pm 0.20$ (for a Salpeter IMF)
provides a similarly good fit so that we can not draw a definitive answer.

Any correlation between a DM quantity and a stellar one may be the
outcome of a hidden interaction between the two galactic
components. In particular, for the Newtonian accelerations,
because of their definition, it is straightforward to show that
$g_{DM}(R_{eff}) = G M_{DM}(R_{eff})/R_{eff}^2 =
M_{DM}(R_{eff})/M_{eff} \times g_{\star}(R_{eff})$ with $M_{eff} =
M_{\star}(R_{eff})$ so that one can look for a correlation between
these two quantities. For the NFW\,+\,Salpeter model, we indeed
get\,:

\begin{displaymath}
\log{g_{DM}(R_{eff})} = 0.21 \log{g_{\star}(R_{eff})} - 7.89 \ .
\end{displaymath}
with $\sigma_{int} = 0.018$, the marginalized constraints being\,:

\begin{displaymath}
\alpha = 0.20_{-0.13}^{+0.13} \ \ , \ \ \beta = -7.90_{-1.28}^{+1.13} \ \ ,
\ \ \sigma_{int} = 0.042_{-0.030}^{+0.043} \ \ .
\end{displaymath}
Actually, we can recast the above relation in a different way. From the
definitions of $g_{DM}(R_{eff})$ and $g_{\star}(R_{eff})$ and the assumption
$\log{g_{DM}(R_{eff})} \propto \alpha \log{g_{\star}(R_{eff})}$, one easily gets\,:

\begin{displaymath}
\log{M_{DM}(R_{eff})} \propto 2(1 - \alpha) \log{R_{eff}} + \alpha \log{M_{eff}}
\end{displaymath}
so that we fit a loglinear relation 

\begin{displaymath}
\log{M_{DM}(R_{eff})} = \alpha_M \log{R_{eff}} + \beta_M \log{M_{eff}} + \gamma_M \ .
\end{displaymath}
For the best fit relation, we find $(1.46, 0.60, 2.79)$
with $\sigma_{int} = 0.006$, while the marginalized constraints (median and $68\%$ CL) read

\begin{displaymath}
\alpha_M = 1.47_{-0.26}^{+0.25} \ \ , \ \ \beta_M = 0.61_{-0.21}^{+0.22} \ \ , 
\end{displaymath}

\begin{displaymath}
\gamma_M = 2.54_{-0.26}^{+0.41} \ \ , \ \ 
\sigma_{int} = 0.029_{-0.020}^{+0.033} \ .
\end{displaymath}
For $\alpha = 0.21$, we expect $\alpha_M \simeq 1.58$ in agreement with our estimate. On the
contrary, $\beta_M$ is significantly larger than $\alpha$ possibly indicating that the ratio
$g_{DM}(R_{eff})/g_{\star}(R_{eff})$ depends on the stellar mass more than expected. However, because of the
correlation between $\beta_M$ and $\sigma_{int}$ induced by the fit, a wrong estimate of the intrinsic scatter
may induce a bias in the best fit $\beta_M$. Since the error bars are quite large, determining $\sigma_{int}$ is
a difficult task as can also be understood noting that the best fit $\sigma_{int}$ is formally outside the
marginalized $68\%$ confidence range (but within the $95\%$ one). Nevertheless, the small value of the rms of the
residuals ($\sigma_{rms} = 0.12$) is strong evidence in favour of a very tight correlation. We also stress that this scaling
relation (although with different coefficients) still holds if we change the IMF or the halo model. Interestingly, this correlation
is pretty similar to the luminosity and mass FP discussed in \cite{Bolton+07} and \cite{HB09}, with the total $M/L$ ratio found
to depend less on the stellar mass density than on the effective radius.

An alternative way to look for the correlation between the stellar and DM mass at $R_{eff}$ may be provided by the DM mass fraction,
$f_{DM}(R_{eff}) = M_{DM}(R_{eff})/[M_{\star}(R_{eff}) + M_{DM}(R_{eff})]$. Since $M_{DM}(R_{eff})$ and $M_{\star}(R_{eff})$ are correlated,
we expect to find a similar correlation between $f_{DM}(R_{eff})$ and mass proxies, such as $M_{\star}$ and $L_V$ (\citealt{Cappellari+06},
\citealt{Bolton+07}, \citealt{HB09}, T09, \citealt{SlacsX}). We indeed find

\begin{displaymath}
\log{f_{DM}(R_{eff})} = 0.49 \log{\left ( \frac{M_{\star}}{10^{11} \ {\rm L_{\odot}}} \right )} - 0.86 \ ,
\end{displaymath}

\begin{displaymath}
\log{f_{DM}(R_{eff})} = 0.51 \log{\left ( \frac{L_V}{10^{11} \ {\rm L_{\odot}}} \right )} - 0.56 \ ,
\end{displaymath}
with $\sigma_{int} = 0.03$ (0.02) and $\sigma_{rms} = 0.13$ (0.12) for the first (second) case. The
marginalized constraints for the $f_{DM}$\,-\,$M_{\star}$ relation are\,:

\begin{displaymath}
\alpha = 0.48_{-0.11}^{+0.11} \ , \ \beta = -0.86_{-0.08}^{+0.08} \ , \ \sigma_{int} = 0.049_{-0.032}^{+0.041} \ ,
\end{displaymath}
while for the $f_{DM}$\,-\,$L_{V}$ we find\,:

\begin{displaymath}
\alpha = 0.51_{-0.09}^{+0.09} \ , \ \beta = -0.57_{-0.01}^{+0.01} \ , \ \sigma_{int} = 0.028_{-0.021}^{+0.038} \ ,
\end{displaymath}
The large error bars on the individual points likely make the estimate of $\sigma_{int}$
biased, but we nevertheless find clear evidence for a DM content
increasing with both $M_{\star}$ and $L_{V}$. Both these correlations are in very good agreement with
what is found in T09 for local ETGs for high luminosity systems ($\log{L_B} \ge 10.4$), notwithstanding the
different models adopted (NFW halo and Sersic stellar profile here vs a full isothermal model in T09). However, the slopes we find are strongly dependent on the model assumptions. Should we use the same NFW model, but the Chabrier
IMF, we get $f_{DM}(R_{eff}) \propto L_V^{0.20}$ and
$f_{DM}(R_{eff}) \propto M_{\star}^{0.16}$, which are fully
consistent with the results we have obtained in \cite{C09}, where
a general galaxy model has been fitted to a subsample of SLACS
lenses\footnote{Note that there is a typo in \cite{C09}, the best
fit relation being $f_{DM}(R_{eff}) \propto M_{\star}^{0.13}$, while the
normalization refers to a Chabrier IMF.}.

\section{Conclusions}\label{sec:sec3}

Much attention has recently been dedicated to investigating
whether some correlations can be found among DM quantities and the
stellar ones with contrasting results pointing towards a universal
DM column density ${\cal{S}}$ (D09, G09) or its variation with
halo mass $M_{200}$ (B09) and luminosity $M_{\star}$ (T09, NRT10, \citealt{A10}).
Here we have addressed this controversy using a sample of
intermediate redshift ETGs using the available data on both the
projected mass within the Einstein radius and the aperture
velocity dispersion to fit four different stellar\,+\,DM halo
models. Motivated by recent findings in the literature
(\citealt{Treu+09}, NRT10) and the analysis of the virial $M/L$
ratio, we have finally chosen a NFW DM halo and a Salpeter IMF
which we have then used as a reference case for investigating the
scaling relations of interest.

Contrary to D09, we find that the column density ${\cal{S}}$,
evaluated at both the halo characteristic radius $r_s$ or the
stellar effective radius $R_{eff}$ is not a universal quantity,
but rather correlates with the luminosity $L_V$ and the stellar
and halo masses $M_{\star}$ and $M_{200}$. Although the slopes of
these correlations depend on the halo model and IMF, assuming our
reference model, the ${\cal{S}}(r_s)$\,-\,$M_{200}$ relation we
find agrees with the B09 one, with a rather similar slope (0.19
vs 0.21), but a smaller zeropoint. As a consequence, our
${\cal{S}}(r_s)$ values are smaller than the B09 ones, but also
smaller than those predicted on the basis of a secondary infall
model (B09, \citealt{dP09}) and $\Lambda CDM$ N\,-\,body
simulations \citep{MDvdB08}. We argue that this discrepancy is
expected considering that these studies do not add a stellar
component to the galaxy model, while here
we take this explicitly into account thus decreasing the DM
content. We have also found that the ensemble average column density in the
central regions is systematically larger than the one in spiral
galaxies in agreement with, e.g., NRT10. This is consistent with
mass accretion in more massive haloes due to merging of
late\,-\,type systems. As an interesting new result, we have shown
that a very tight loglinear relation among $M_{DM}(R_{eff})$,
$R_{eff}$ and $M_{eff}$ can be found leading to a DM mass fraction
which positively correlates with both the stellar luminosity and
mass (\citealt{Cappellari+06}, \citealt{Bolton+07},
\citealt{HB09}, T09).

The limited mass and luminosity range probed and the large errors on
the different quantities involved prevent us from drawing a definitive
answer on the slope and normalization of the above scaling relations.
Moreover, a larger dataset should also allow us to make a more detailed
investigation of the impact of halo profiles and IMF assumptions. Should
these further tests confirm our results, DM scaling relations can provide
a valuable tool in understanding the physical processes which drive galaxy
formation and evolution.

\section*{Acknowledgments}

We warmly thank the anonymous referee for his/her suggestions and
the help to improve the paper and Garry Angus for a careful reading of the manuscript. 
CT was supported by the Swiss National Science Foundation.

\end{document}